# Self-optimizing layered hydrogen evolution catalyst with high basal-plane activity

Yuanyue Liu[1+], Jingjie Wu[1+], Ken P. Hackenberg[1+],Jing Zhang[1], Y. Morris Wang[2], Yingchao Yang[1], Kunttal Keyshar[1], Jing Gu[3], Tadashi Ogitsu[2], Robert Vajtai[1], Jun Lou[1], Pulickel M. Ajayan[1], Brandon C. Wood*[2], Boris I. Yakobson*[1]

[1]Department of Materials Science and Nano-Engineering, Rice University, Houston, TX 77005.

[2]Lawrence Livermore National Laboratory, Livermore, CA 94550.

[3]National Renewable Energy Laboratory, Golden, CO, 80401

[+]These authors contribute equally

*Correspondence to: brandonwood@llnl.gov, biy@rice.edu

**Hydrogen is a promising energy carrier and key agent for many industrial chemical processes[1]. One method for generating hydrogen sustainably is via the hydrogen evolution reaction (HER), in which electrochemical reduction of protons is mediated by an appropriate catalyst—traditionally, an expensive platinum-group metal. Scalable production requires catalyst alternatives that can lower materials or processing costs while retaining the highest possible activity. Strategies have included dilute alloying of Pt[2] or employing less expensive transition metal alloys, compounds or heterostructures (e.g., NiMo, metal phosphides, pyrite sulfides, encapsulated metal nanoparticles)[3-5]. Recently, low-cost, layered transition-metal dichalcogenides ($MX_2$)[6] based on molybdenum and tungsten have attracted substantial interest as alternative HER catalysts[7-11]. These materials have high intrinsic per-site HER activity; however, a significant challenge is the limited density of active sites, which are concentrated at the layer edges.[8,10,11]. Here we use theory to unravel electronic factors underlying catalytic activity on $MX_2$ surfaces, and leverage the understanding to report group-5 $MX_2$ (*H*-$TaS_2$ and *H*-$NbS_2$) electrocatalysts whose performance instead derives from highly active basal-plane sites. Beyond excellent catalytic activity, they are found to exhibit an unusual ability to optimize their morphology for enhanced charge transfer and accessibility of active sites as the HER proceeds. This leads to long cycle life and practical advantages for scalable processing. The resulting performance is comparable to Pt and exceeds all reported $MX_2$ candidates.**

The HER proceeds via two steps: (i) H adsorbs on the catalyst by $H^+ + e^- + * \rightarrow H*$ (Volmer reaction), where * denotes a catalytic site; (ii) $H_2$ is formed and desorbed by either $2H* \rightarrow H_2 + 2*$ (Tafel reaction) or $H^+ + e^- + H* \rightarrow H_2 + *$ (Heyrovsky reaction).[12] Among the major factors[13] that determine the HER rate is the balance between adsorption and desorption—an empirical rule known as the Sabatier principle, typified by the "volcano plot" [2,10,12-16]. If the substrate interaction is too weak, then the Volmer reaction is inhibited; if it is too strong, then the Tafel/Heyrovksy reaction cannot proceed. The relative adsorption free energy of the H* intermediate therefore acts as an indicator of the catalytic activity, and has been widely used to evaluate catalyst candidates [2,10,16].

For a deeper understanding of adsorption behavior on $MX_2$ (M = transition metal; X = S, Se, Te), we examine how the underlying electronic structure is modified by the presence of the H* intermediate. We find that dilute H adsorption leaves the profile of the electronic density of states (DOS) largely intact, with complete charge transfer from the adsorbate to the substrate. Consequently, its dominant effect in both metal and semiconductor $MX_2$ is to populate states at or near the lowest unoccupied state ($\varepsilon_{LUS}$)—the conduction band minimum for semiconductors or the Fermi level for metals. The general behavior is illustrated schematically in Fig. 1a, and implies that $\varepsilon_{LUS}$ is the key determinant of adsorption strength on $MX_2$ surfaces (full DOS calculations and charge densities for specific $MX_2$ candidates, along with descriptions of underlying physical mechanisms, can be found in Supplementary information Fig. S1). We point out that this same $\varepsilon_{LUS}$ descriptor has been recently shown to predict lithium adsorption on carbon[17], and parallels relationships observed for chlorine evolution[18,19] and oxygen reduction[20] on certain oxides. Moreover, for metallic systems, $\varepsilon_{LUS}$ is closely connected to the work function, which has long been shown to correlate with HER activity on elemental metals[14]. We verified the direct correlation between $\varepsilon_{LUS}$ and H* adsorption energy ($E_a$) on the basal plane for a training set of known $MX_2$ at a dilute concentration (see Methods and Fig. S1). As shown in Fig. S2, other probable descriptors do not exhibit the same level of correlation; in particular, the breakdown of the d-band center rule likely owes to the additional contributions from X p states near the Fermi level). We therefore adopt $\varepsilon_{LUS}$ as a descriptor for selecting basal-plane-active $MX_2$ catalysts based on intrinsic properties. This also renders explicit evaluation of H* adsorption unnecessary, offering a computational advantage for more efficient screening.

In Fig. 1B, we apply our $\varepsilon_{LUS}$ criterion to all $MX_2$ substrates. We consider the most stable phases (*H* for group 5 and 6, *T* for group 4 and 10, *T'* for group 7; structures are shown in Fig. S2). We select as a target criterion for viable candidates -6.4 eV < $\varepsilon_{LUS}$ < -5.5 eV, which corresponds to -0.5 eV/H < $E_a$ < +0.5 eV/H based on Fig. S1 (this window accounts for additional contributions to the free energy). Note that *H*-$MoX_2$ and *H*-$WX_2$ monolayers exhibit comparatively high $\varepsilon_{LUS}$ (> -4.5 eV), which leads to weak adsorption that inhibits the Volmer reaction and prevents basal-plane activity in these materials. Two general features are observed: (i) for a given M, $\varepsilon_{LUS}$ (and hence $E_a$) increases in the order S < Se < Te; and (ii) metallic $MX_2$ candidates (from groups 4 and 5) have lower $\varepsilon_{LUS}$ and hence stronger $E_a$ than semiconducting $MX_2$ candidates (from groups 6, 7, and 10). Among the viable candidates, the group 5 metal disulfides (*H*-$VS_2$, *H*-$NbS_2$, and *H*-$TaS_2$) are clearly the most promising, having a low $\varepsilon_{LUS}$ (< -5.8 eV) near the center of our window of interest. Explicitly accounting for additional contributions to the free energy of H* adsorption provides further confirmation that these three materials should have high activity per basal surface site (see Methods and Fig. S3). We point out that our calculations demonstrate a thermodynamic preference for more dilute H adsorption at zero overpotential (Fig. S3); this is different from Pt, which favors a higher coverage at zero overpotential. Nevertheless, we predict that high per-site activity will be retained as the overpotential is increased to access higher coverages. The potential of the group 5 metal disulfides for HER catalysis has been noted in other recent theoretical analyses by explicit evaluation of H adsorption.[21,22]

We successfully synthesized and tested *H*-$TaS_2$ and *H*-$NbS_2$ (see Methods and Fig. S4 for synthesis and characterization details). As shown in Fig. 2, indeed both *H*-$TaS_2$ and *H*-$NbS_2$ demonstrate extraordinary HER catalytic performance. Although we were not able to synthesize *H*-$VS_2$ at this stage, *T*-$VS_2$ has also been found to have intrinsic activity[23], in agreement with its relatively low $\varepsilon_{LUS}$ (~ -5.7 eV). Figure 2a displays the polarization curves for HER electrocatalysis; for comparison under identical conditions, results for $MoS_2$ and Pt (see Fig. S11 for Pt) are also shown. After 5000 cycles, both $H$-$TaS_2$ and $H$-$NbS_2$ exhibit an extremely low onset potential. A current density of 10 $mA/cm^2$ (a standard for comparison[3]) is reached at only -60 mV versus the reversible hydrogen electrode (RHE). As shown in Fig. 2b, the materials also have very low Tafel slopes (37 mV/dec for $H$-$TaS_2$; 30 mV/dec for $H$-$NbS_2$) and high exchange current densities ($7.8\times10^{-2}$ $mA/cm^2$ for $H$-$TaS_2$; $1.2\times10^{-2}$ $mA/cm^2$ for $H$-$NbS_2$) even with small loading (10-55 $\mu g/cm^2$). These performance characteristics are far superior to other $MX_2$ materials tested in this work or reported elsewhere (see SI Table S1), and rank both materials among the best non-noble-metal catalysts currently available[3]. The relevance of $MX_2$ crystal structure in determining catalytic activity[24-26] was also demonstrated, as Fig. 2a shows that the performance of a synthesized sample of the also-metallic $T$-$TaS_2$ is far inferior to the ground-state $H$ phase [27,28] (this is consistent with calculations in Fig. S1 showing $T$-$TaS_2$ has a much higher $\varepsilon_{LUS}$). This contrasts with $T$-$MoS_2$, which exhibits higher activity than edge-active $H$-$MoS_2$ [25,26].

The $H$-$TaS_2$ and $H$-$NbS_2$ multilayer platelets exhibit an additional unusual benefit, in that repeated catalysis of hydrogen evolution results in continual and dramatic improvement in catalytic performance before reaching steady state (for $H$-$TaS_2$, see Fig. 2c and Fig. S5, as well as the potentiostatic measurements in Fig. S6; for $H$-$NbS_2$, see Fig. S5). This self-optimizing behavior has extraordinary practical advantages compared to more complex approaches for optimizing $MX_2$ catalysts, in that it enables highly scalable processing with minimal additional treatment. Microscopy analysis indicates that the performance enhancements are associated with a morphological evolution of the catalyst. In particular, comparing atomic force microscopy (AFM), transmission electron microscopy (TEM) and scanning electron microscopy (SEM) results before and after cycling (Fig. 3a-d and Fig. S7) for $H$-$TaS_2$ illustrates that the platelets become thinner, smaller, and more dispersed. By contrast, there are no discernible changes in the local crystal structure or chemical composition of $H$-$TaS_2$, as confirmed by Raman spectra, energy dispersive spectroscopy (EDS), X-ray photoelectron spectroscopy (XPS) and high-resolution TEM (Fig. S8). Analogous self-optimizing morphology changes and chemical intactness are also observed for $H$-$NbS_2$ (see Fig. S7-8).

The changes in the catalyst morphology upon cycling have two key beneficial consequences for catalytic activity, which are illustrated in Fig. 3e-f. The first consequence is to shorten the interlayer electron-transfer pathways due to sample thinning. This is particularly beneficial for weakly bound layered materials, which tend to have poor electron transport along the stacking direction (Methods and Fig. S4). Indeed the electrochemical impedance spectra (EIS; see Fig. S5) show that a key change upon cycling is connected to decreased charge-transfer resistance (Fig. 3e for $H$-$TaS_2$ and Fig. S5 for $H$-$NaS_2$; also see Methods for details of the equivalent circuit modeling), which translates to improved electrical conductivity in the absence of additional changes to the local chemistry. Similar conclusions can be drawn by analyzing the decrease in the Tafel slope upon cycling (Fig. 2c inset for $H$-$TaS_2$; Fig. S5 for $H$-$NbS_2$), which signals a change in the rate-determining step away from the initial electron transfer (Volmer) process as electrical conductivity improves and charge transfer becomes more facile[29]. The second consequence is to increase the effective active surface area by improving accessibility of aqueous protons to basal-plane sites. This is evidenced in the increase of the effective double-layer capacitance (Fig. 3e for $H$-$TaS_2$; Fig. S5 for $H$-$NbS_2$), which is expected to scale roughly with the electrolyte-accessible surface area as the material is cycled (see Methods for calculation details and assumptions). Note that the very large magnitude of the capacitance increase indicates that a significant fraction of the newly accessible surface area arises from additional interior sites connected to higher porosity; these interior sites are also likely to have shorter electron transfer pathways, offering a secondary benefit. In summary, both electron transport and accessible surface area are enhanced by the morphology changes, acting in concert to boost overall catalytic performance. Moreover, the current scales strongly with the solvent-accessible surface area upon cycling (Fig. S5), further supporting our conclusion that active basal-plane sites are key to the observed performance (although the effective surface area is difficult to assess quantitatively[30], a qualitative increase of the surface area with cycling is nonetheless evident) Indeed, the extraordinary performance is best explained by considering both high per-site activity and a large number of active sites derived from basal-plane activity..

Moreover, inductively coupled plasma mass spectrometry (ICP-MS), CO stripping voltammetry (see Methods and Fig. S9) and XPS survey scan (Fig. S10) measurements rule out the possibility that the high catalytic activity might instead arise from extrinsic contamination (e.g., by Pt), confirming that the basal-plane activity is indeed intrinsic.

We propose that the high basal-plane activity is directly responsible for the self-optimizing morphological changes. Because $H$-$TaS_2$ and $H$-$NbS_2$ are weakly bound layered materials, $H_2$ produced at basal-plane sites between layers becomes trapped and perforates or peels away layers in order to escape, resulting in a thinner and more porous catalyst (Fig. 3f). This mechanism is analogous to reports of lithium intercalation and reaction with water in $MX_2$, where hydrogen gas produced between layers can cause exfoliation[25]. It is also consistent with the observation that graphene can be delaminated from metal substrate by electrochemical $H_2$ bubbling[31,32]. In fact, electrochemical H intercalation and exfoliation of $H$-$TaS_2$ have been reported in early literature (although a different exfoliation mechanism is suggested).[33,34] Reliance on basal-plane activity would also explain why these same improvements are not seen in edge-active $H$-$MoS_2$ (Fig. 2c).

In summary, we present highly basal-plane-active $MX_2$ electrocatalysts for HER, based on theoretical prediction and experimental validation. The success was facilitated by a fundamental understanding of underlying electronic motivations, from which an appropriate descriptor was devised. The catalysts are shown to have exceptional per-site catalytic activity and a high number of active sites compared to edge site-limited counterparts. We also find that they exhibit unusual self-optimizing performance as they catalyze hydrogen evolution, which derives from beneficial morphological changes that enhance charge transfer and accessibility of active sites. As a result, high performance can be achieved with minimal catalyst loading and processing, offering significant practical advantages for scalability and cyclability. We point out that performance might be further optimized by applying chemical and engineering strategies demonstrated for other $MX_2$ materials[11]. Our work opens the door to the use of this novel and practical type of catalyst, and lays out a compelling scheme for assessing activity in similar classes of materials.

## Methods

Methods and any associated references are available in the online version of the paper.

**Acknowledgments:**

We thank John Turner (NREL), Wu Zhou (ORNL), Woon Ih Choi (LLNL), Aditya Mohite and Gautum Gupta (LANL) for valuable discussions. BW and YL acknowledge funding from LLNL LDRD Grant 12-ERD-053, with computing support from the LLNL Institutional Computing Grand Challenge program. TO and BW acknowledge additional support from the U.S. Department of Energy Fuel Cell Technologies Office. A portion of this work was performed under the auspices of the U.S. Department of Energy by LLNL under Contract DE-AC52-07NA27344. KH acknowledges funding from PIRE-2 Grant OISE-0968405. JW and KK acknowledge funding from MURI 2D Grant W911NF-11-1-0362. YY, JZ and JL acknowledge support from the Welch Foundation grant C-1716.

**Author Contributions**:

YL conceived the idea and performed the theory calculations with guidance from TO, BW and BY. KH synthesized the samples, JW performed the electrochemical testing. JW and KH performed a majority of the materials characterization, under the guidance of RV, JL and PA. Other authors provided additional sample characterization.

**Additional information:**

Supplementary information is available in the online version of the paper. Reprints and permissions information is available online at www.nature.com/reprints. Correspondence and requests for materials should be addressed to BW and BY.

**Competing financial interests**

The authors declare no competing financial interests

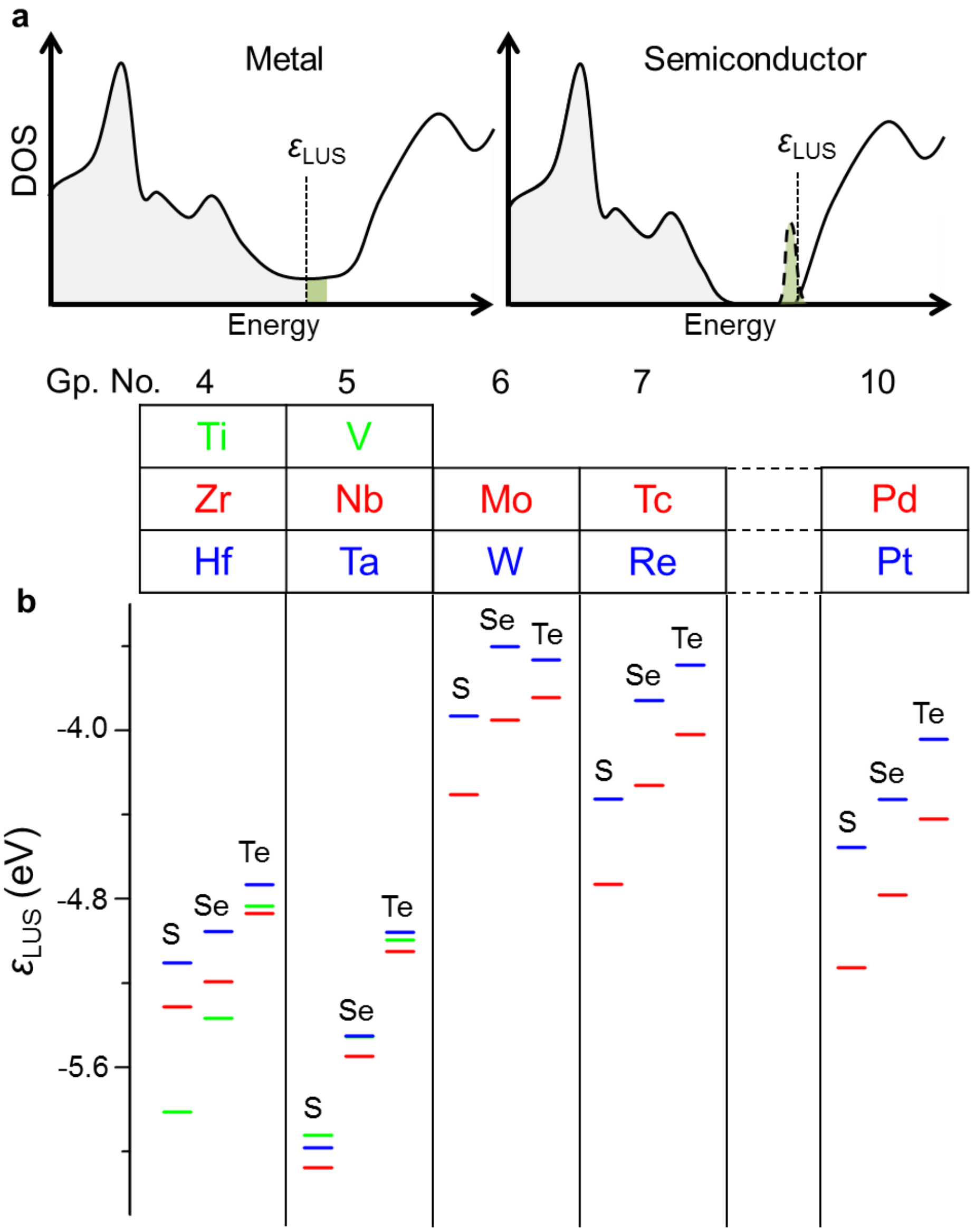

**Figure 1| Electronic origin of $MX_2$ surface activity and the derived descriptor ($\varepsilon_{LUS}$) for catalysts screening. (a)** Schematic of the $MX_2$ DOS, showing initial filled (gray) and empty states, as well as newly filled (green) states upon dilute H adsorption. In metals, the Fermi level is slightly elevated, whereas in semiconductors, a shallow state appears near the conduction band edge; in each case, the newly occupied states closely follow $\varepsilon_{LUS}$. **(b)** Computed $\varepsilon_{LUS}$ for all $MX_2$ candidates. Row 4/5/6 elements are shown in green/red/blue, with the different chalcogens separated into columns within each group.

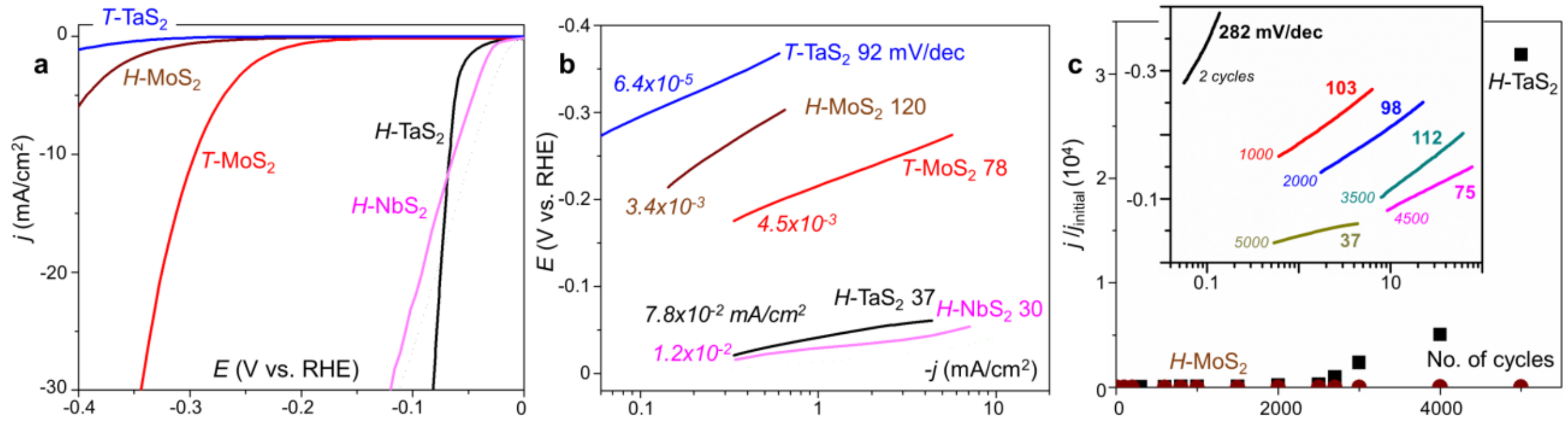

**Figure 2| HER electrocatalysis on *H*-$TaS_2$ and *H*-$NbS_2$.** (**a**) Polarization curves (iR-corrected) for *H*-$TaS_2$, *H*-$NbS_2$, *H*-$MoS_2$, *T*-$MoS_2$, and *T*-$TaS_2$ measured in Ar bubbled 0.5 M $H_2SO_4$ with a scan rate of 5 mV $s^{-1}$. *H*-$TaS_2$, *H*-$NbS_2$ and *H*-$MoS_2$ were initially cycled for 5000 cycles between 0.2 and -0.6 V vs. RHE at 100 mV $s^{-1}$. (**b**) Corresponding Tafel plots for catalysts in (a); Tafel slopes (mV/dec) and exchange current densities (mA/$cm^2$) are shown. (**c**) Comparison of current density increase during cycling of *H*-$TaS_2$ versus *H*-$MoS_2$ (recorded at -0.1 V for *H*-$TaS_2$; *H*-$MoS_2$ is recorded at -0.3 V to surmount low activity at -0.1 V). Inset shows Tafel plots of *H*-$TaS_2$ measured at different cycles.

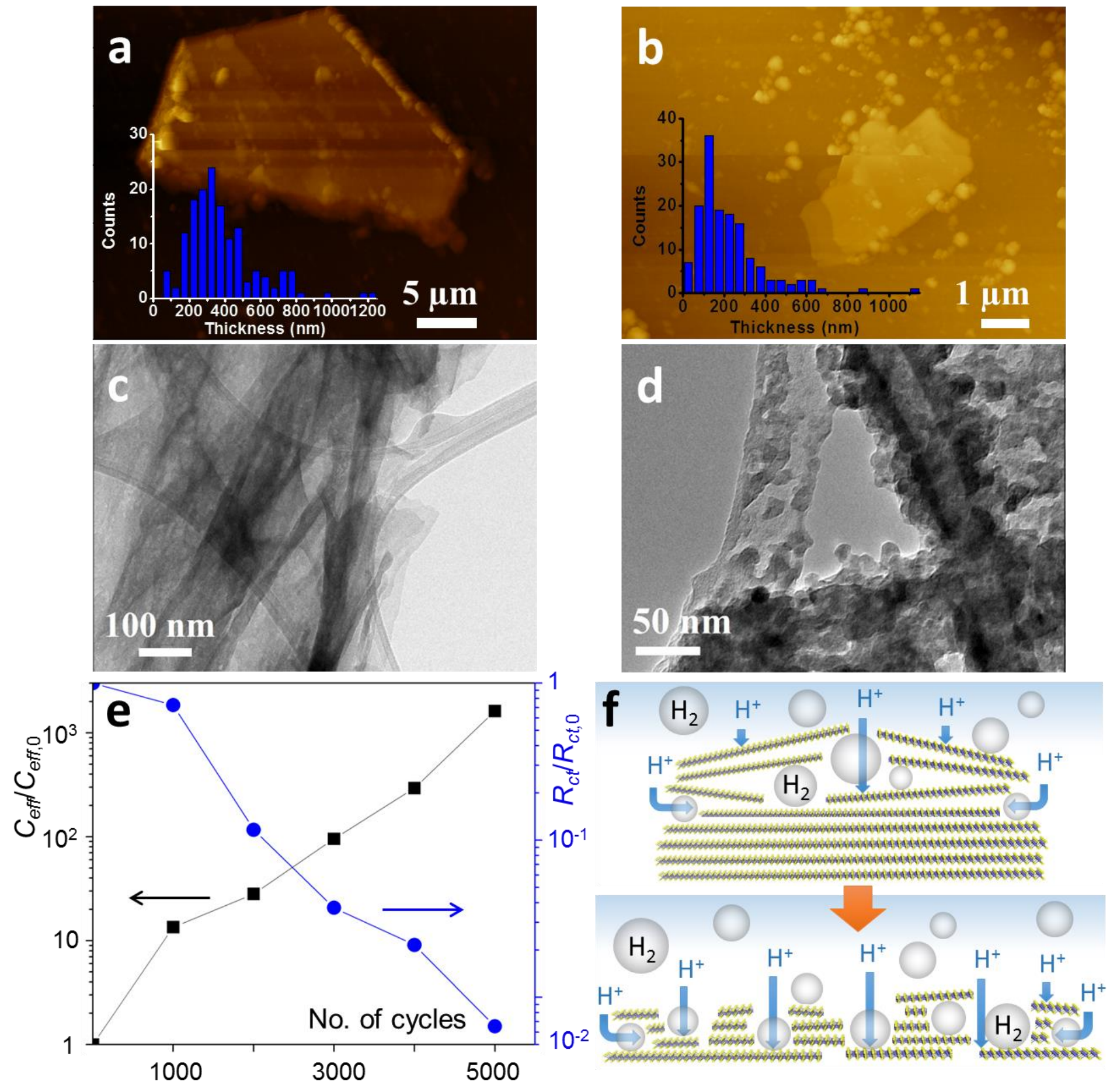

**Figure 3| Origin of self-optimizing behavior.** AFM (a-b) and TEM (c-d) images of $H$-$TaS_2$ before (a and c) and after (b and d) cycling. Insets in (a-b) show the statistical distributions of thicknesses. (e) Change of effective capacitance ($C_{eff}/C_{eff,0}$ where 0 denotes the initial value) and charge-transfer resistance ($R_{ct}/R_{ct,0}$) as a function of cycles for a different $H$-$TaS_2$ sample. Both quantities are extracted from the EIS (see Methods and Fig. S5). (f) Schematics of the proposed mechanism for the morphology change, in which hydrogen evolution at basal-plane sites of stacked layers causes perforation and exfoliation.

**Computational details:** Spin-polarized DFT calculations were performed using Projector Augmented Wave (PAW) pseudopotentials[1,2] and the Perdew-Burke-Ernzerhof (PBE) exchange-correlation functional,[3] as implemented in VASP.[4,5] All structures were based on $MX_2$ monolayers, and were relaxed until the force on each atom is less than 0.01 eV/Å. Vacuum space in the direction perpendicular to the basal plane was kept to >15 Å. The adsorption energy $E_a$ is defined as

$$E_a = E(H+MX_2) - E(MX_2) - E(H_2)/2 \quad (1)$$

where $E(H+MX_2)$, $E(MX_2)$ and $E(H_2)$ are the energies of H-adsorbed $MX_2$, pure $MX_2$, and an $H_2$ molecule, respectively. For $MX_2$ in the *H* or *T* phase, $E_a$ was calculated using a 4x4 supercell, while for the *T'* phase, we used a 4x2√3 cell. $\varepsilon_{LUS}$ was calculated based on the primitive cell. For the HER at pH = 0 and at zero potential relative to the standard hydrogen electrode, the free energy of $H^+ + e^-$ is by definition the same as that of 1/2 $H_2$ at standard conditions. Using this value as a zero reference, we estimate the free energy as

$$G_{tot} = (E_a + E_{ZP} - TS + E_{solv})*n_H \quad (2)$$

where $E_{ZP}$ is the zero-point energy, $TS$ is the entropy contribution, and $E_{solv}$ is the solvation energy. $E_{solv}$ was evaluated using the VASPsol implementation of the implicit solvation model of Ref. [6]. Coverage-dependent values of $G_{tot}$ for *H*-$TaS_2$, *H*-$VS_2$, and *H*-$NbS_2$ can be found in Fig. S3. The differential free energy $G_{diff}$ at the equilibrium H coverage represents the free energy cost to adsorb/desorb H on/from the catalyst, which in turn reflects the kinetics of catalysis near equilibrium [7-10]:

$$G_{diff} = \partial G_{tot}/\partial n_H \quad (3)$$

Calculations of $G_{diff}$ were performed at H:M = 1:16 for $MX_2$ surfaces, H:M = 1:2 for $MoS_2$ edges[8], and H:M = 1:1 for M=Pt, Ni (the latter case was based on calculations of $G_{tot}$ for Pt and Ni that showed equilibrium monolayer surface coverages). Corresponding plots can be found in Fig. S3. We have also assessed the possible effect of interface polarization due to an applied potential by including a large electric field (~ 0.7 V/nm) pointing towards the substrate. Even with such a high field strength, the effect on $G_{diff}$ is negligible, and our conclusion of high surface activity for group 5 $MX_2$ remains robust.

**Synthesis:** *H*-$TaS_2$ and *H*-$NbS_2$ crystal platelets were grown by chemical vapor deposition on $SiO_2$/Si substrates in a three-stage furnace. *H*-$TaS_2$ was derived from sulfur and tantalum chloride powders and gaseous hydrogen precursors via the following reactions:

$H_2 + 1/8S_8 \rightarrow H_2S$

$TaCl_5 + 2H_2S + 1/2H_2 \rightarrow TaS_2 + 5HCl$

The sulfur, tantalum chloride, and growth substrate regions were held respectively at ~250 ºC, ~300 ºC, and ~750 ºC for a 10 minute growth period with a 20 sccm flow of Ar/$H_2$ (85:15). *H*-$TaS_2$ platelets were converted to the *T* phase by heating in a S/Ar atmosphere at 900 ºC for 1 hour and then rapidly quenching. *H*-$NbS_2$ crystal platelets were grown in the similar way to *H*-$TaS_2$ except that $NbCl_5$ was used as Nb element precursor and the substrate was held at 550 ºC. Specifically, the $NbCl_5$ powder and S powder were placed in two different zones with temperatures of 250 °C and 300 °C, respectively, while the $SiO_2$/Si substrate was located in a third zone downstream with the temperature held at 550 °C. The carrier gas (Ar/$H_2$ 85:15 by volume) and flow rate (20 sccm) were the same as those for *H*-$TaS_2$ growth. Multiple sample batches were prepared and tested.

*T*-$MoS_2$ nanosheets were prepared by *n*-butyl lithium insertion and reaction with water procedure, from commercial *H*-$MoS_2$ plates (Sigma-Aldrich)[11,12].

**Material characterization:** SEM images and EDS were recorded on an FEI Quanta 400 microscope. XPS spectra were collected using a PHI Quantera X-ray photoelectron spectrometer. AFM measurements were taken using an Agilent Picoscan5500 AFM equipped with a silicon tapping mode tip (AppNano). In the case of comparing the morphology before and after potential cycling, SEM and AFM images were taken on the samples loaded onto the glassy carbon plate. TEM images were collected on a JEOL 2100F TEM. Samples were prepared by drop-drying a diluted suspension in isoproponal onto copper grids covered with lacy carbon films. X-ray diffraction (XRD) was carried out on a Rigaku D/Max Ultima II Powder XRD. Our CVD-synthesized NbS2 is not dense enough for powder XRD measurements , leading to an unreliably low signal-to-noise ratio. Raman spectra were taken at an excitation wavelength of 514 nm. The Raman spectra show that both $TaS_2$ and $NbS_2$ are 2H-phase[13] (here 2 means the unit cell of bulk material has two layers, and H indicates a hexagonal atomic patterning in each layer). For comparison, the 1T phase reports peaks at ~285, ~417, and ~607[14]; and the 3R phase reports peaks at 290±5, 330±3, 386±2, and 458±3[13], neither of which is consistent with our data. The in-plane and out-of-plane resistivity of *H*-$TaS_2$ and *H*-$MoS_2$ were measured by the I-V curves and the devices were fabricated through assistance of e-beam lithography.

**Electrode preparation:** The materials were transferred from the $SiO_2$/Si wafers by a PMMA transfer method. First, PMMA was spin coated onto the wafers. The PMMA covered wafer was etched in KOH solution, which removed the $SiO_2$ layer. Next, the PMMA was dissolved in acetone. Finally, the catalyst materials were isolated by centrifuging at 10000 rpm for 5 min followed by drying in $N_2$ atmosphere. A catalyst ink was made by mixing the obtained $TaS_2$/$NbS_2$, isopropanol, deionized water and Nafion (0.5 wt. %) and sonicating for 30 min. The catalyst ink was then dropped onto a glassy carbon electrode, which served as the working electrode. The resulting catalyst loadings for *H*-$TaS_2$ and *H*-$NbS_2$ electrodes were ~55 and ~10 $\mu g/cm^2$, respectively. The catalyst loadings for all electrodes are listed in Table S1.

**Electrochemical performance:** Electrochemical measurements were performed in a three-electrode electrochemical cell using an Autolab PGSTAT302N potentiostat. All measurements were performed in 50 mL of 0.5 M $H_2SO_4$ (aq) electrolyte (pH = 0.16) prepared using 18 MΩ deionized water purged with Ar gas (99.999%). The glassy carbon electrode (CH Instruments, Dia. 3 mm) casted by the samples was employed as the working electrode and a saturated calomel electrode (SCE) (CH Instruments) was used as a reference electrode. For *H*-$NbS_2$ electrode, a graphite rod was used as a counter electrode. For *H*-$TaS_2$ electrode, either a graphite rod or Pt foil was used as counter electrode in the multiple samples measurement. The self-improving performance of *H*-$TaS_2$ was observed in both counter electrodes. A glassy carbon plate loaded with *H*-$TaS_2$ or *H*-$NbS_2$ samples was also employed as a working electrode in order to monitor the morphology change during potential cycling. As a comparison, carbon supported Pt (Pt/C, 20%, Alfa Asaer) was also tested under identical condition with a Pt loading of 25 μg/cm$^2$. The electrolyte was stirred through use of a magnetic stir bar during the electrochemical test to improve the mass transport. The SCE was calibrated in the high purity $H_2$ saturated electrolyte using platinum as both working and counter electrode. Cyclic voltammetry was run at a scan rate of 1 mV s$^{-1}$, and the average of the two potentials at which the current crossed zero was taken to be the thermodynamic potential for the HER. All reported potentials are referenced to RHE. In 0.5 M $H_2SO_4$, $E$ (RHE) = $E$ (SCE) + 0.254 V. HER activity was measured using linear sweep voltammetry between +0.10 ~ -0.50 V vs. RHE with a scan rate of 5 mV s$^{-1}$. The stability was evaluated by the potential cycling performed using cyclic voltammetry initiating at +0.2 V and ending at -0.6 V vs. RHE at either 100 mV s$^{-1}$ or 5 mV s$^{-1}$. All data are corrected for a small ohmic drop measured by EIS.

**Additional characterization:** EIS was performed at a bias potential of -0.1 V vs. RHE while sweeping the frequency from 1 MHz to 10 mHz with a 5 mV AC amplitude. The EIS data were fitted to a Randles equivalent circuit consisting of an ohmic resistance $R_{ohm}$ in series with a charge-transfer resistance ($R_{ct}$)/constant-phase element (CPE) parallel combination. $R_{ct}$ was also confirmed to scale with the exponential of the overpotential, further verifying its assignment as charge-transfer resistance. The power $n$ of the CPE was found to fall within the range 0.68-0.78, indicating frequency dispersion likely connected to porosity, surface roughness, or diffusion factors. The effective double-layer capacitance $C_{eff}$ was computed as $C_{eff} = 1/R_{ct} * (QR_{ct})^{1/n}$, where $Q$ is the CPE coefficient. This formula is intended to partially account for the CPE frequency dispersion. Although the low values of $n$ make precise determination of the electrolyte-accessible catalyst surface area difficult, the qualitative scaling behavior with cycling can be safely assessed given the magnitude of the associated increases. For example, after 5000 cycles, ($1/R_{ct} * R_{ct}^{1/n}$) for $H$-$TaS_2$ changes by a factor of 3.6, compared to ~1800 for $Q$, confirming that the scaling of the capacitance is not a byproduct of decreased $R_{ct}$ but rather is chiefly associated with increased surface area. Scaling plots of current versus $C_{eff}$ (normalized by the initial values) for $H$-$TaS_2$ can be found in Fig. S5.

The anisotropy of the electrical resistance was confirmed by direct resistivity measurements on pre-cycled samples of $H$-$TaS_2$ (see Fig. S4), which yielded 8 Ωcm for out-of-plane resistivity compared with 3 x 10$^{-5}$ Ω cm for in-plane resistivity (for comparison, $H$-$MoS_2$ demonstrated much larger in-plane resistivity of 1.7 Ω cm).

To measure the Faradaic efficiency and confirm $H_2$ as the reaction product, a gas-tight electrochemical setup with two burettes filled with electrolyte solution (one for $H_2$ collection and one for $O_2$ collection) was applied to collect and periodically measure the increased gas volume due to $H_2$ generation in the HER compartment by $H$-$TaS_2$. The collected gas was further confirmed by gas chromatography (Shimadzu GC-2010 Plus) with a Carbxen™ 1010 PLOT column and a thermal conductivity detector by direct injection. As shown in Fig. S9, the periodical measurement of $H_2$ production rate matches well with the theoretical value calculated from the charges passed under -0.4 V vs. RHE (not iR corrected). $H_2$ production for the first 5 h during the activation process continuously increased up to ~6.9 ml (0.31 mmol, theoretical yield should be 7.7 ml or 0.34 mmol), which account for ~90% of Faradaic Efficiency (FE). A FE of ~94% was calculated based on the collection of gas (38 ml, theoretical yield should be 40.4 ml or 1.8 mmol) in the subsequent 19 h electrolysis.

**Tests for Pt contamination:** Inductively coupled plasma mass spectrometry (ICP-MS) was conducted on a Thermo Scientific ICAP Q instrument with CETAC ASX-520 auto sampler. A platinum standard was prepared from Inorganic Ventures MS Pt-10 ppm. A glassy carbon electrode with the *H*-$TaS_2$ catalyst after 5000 cycles was digested in aqua regia solution (EMD OmniTrace hydrochloric acid and nitric acid = 3:1 by volume) overnight. The resulting solution was diluted into 3.75% HCl, 1.25% $HNO_3$ for analysis, and the experiments were repeated three times. Comparable trace amounts of Pt were detected in the diluted $TaS_2$ sample as in the aqua regia solvent solution (0.004 – 0.005 ppb, compared to the detection limit of 0.003 ppb), indicating *H*-$TaS_2$ electrode is free of Pt contamination. Because CO easily and strongly adsorbs on the surface of Pt, CO stripping voltammetry was also used as a secondary quantitative measurement of any trace amount of Pt on the electrode. After 5000 cycles, the CO stripping voltammogram for *H*-$TaS_2$ had exactly the same shape with the background and no CO oxidation peak was observed indicating no contamination of the long-term cycled *H*-$TaS_2$ electrode.

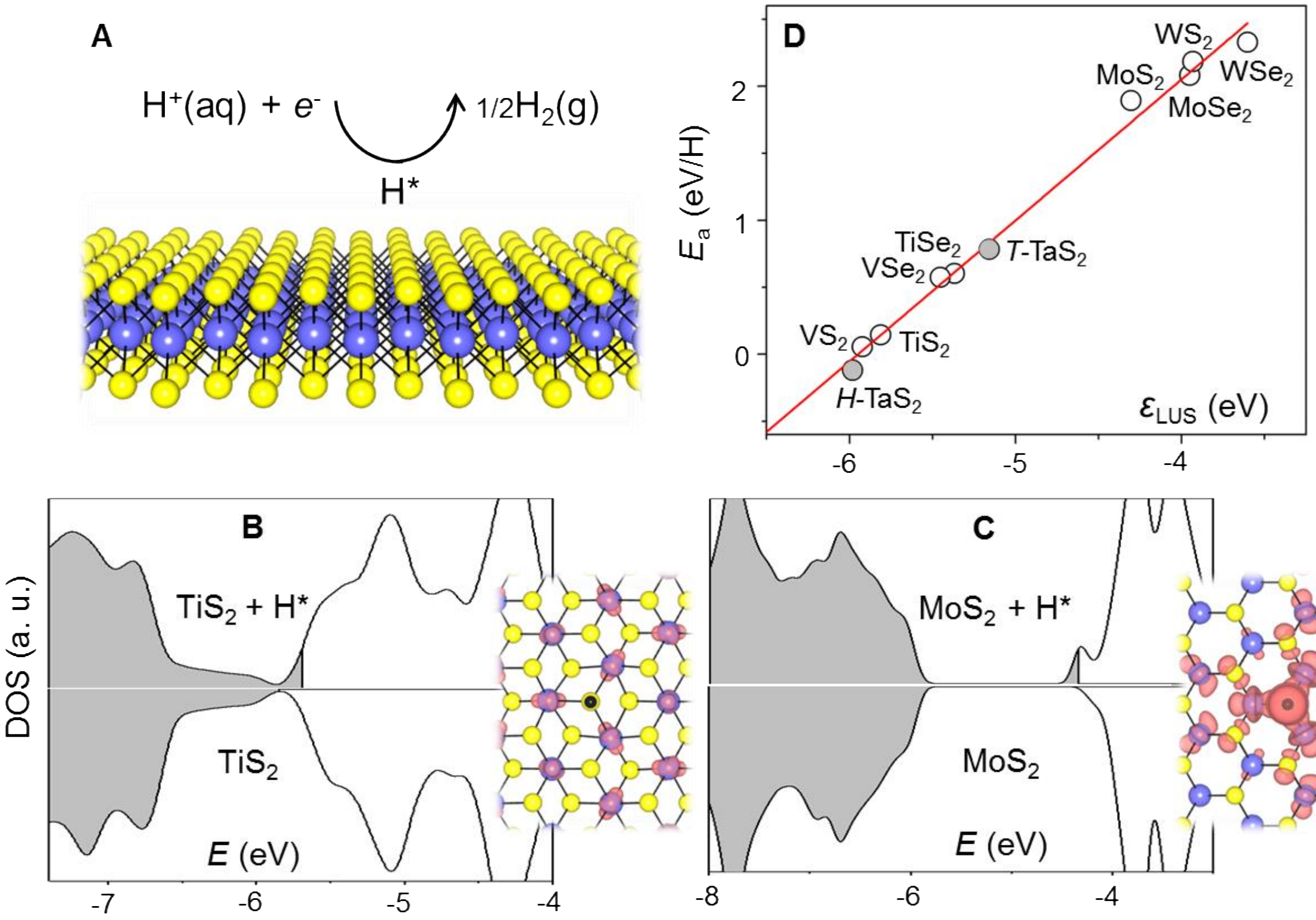

**Figure S1 | $MX_2$ surface activity and its electronic origin. (A)** Schematic of $MX_2$ surface catalyzed HER. M: blue; X: yellow. (**B**) and (**C**) Electronic densities of states (DOS) of representative pristine and H-adsorbed $MX_2$ systems. DOS plots represent a coverage of H:M = 1:16, with the filled states shown in gray (energies are referenced to vacuum for the pristine case). Left/right panels represent metallic (here, $TiS_2$) and semiconducting (here, $MoS_2$) variants. Insets show charge density isosurfaces for states within the energy range of the Fermi level to -0.025 eV below. H: black; charge density isosurface: red. For the metallic system (left panel), H

adsorption does not change the overall DOS profile, instead shifting the Fermi level slightly to reflect complete charge transfer from the H adsorbate. The charge density distribution shows that the transferred electrons are delocalized throughout the M layer. For the semiconducting system (right panel), the DOS profile also remains largely intact, and the H occupies a very shallow *n*-type dopant level. The charge density distribution shows that this state is quasi-localized in space. In the dilute adsorption limit, both cases are well described by $\varepsilon_{\mathrm{LUS}}$. (**D**) Correlation between the $\varepsilon_{\mathrm{LUS}}$ descriptor and the surface adsorption energy $E_a$ (Eq. 1, Methods; negative indicates stronger binding) of H for a test series of $MX_2$ candidates. *H*- and *T*-$TaS_2$ are shown as filled circles.

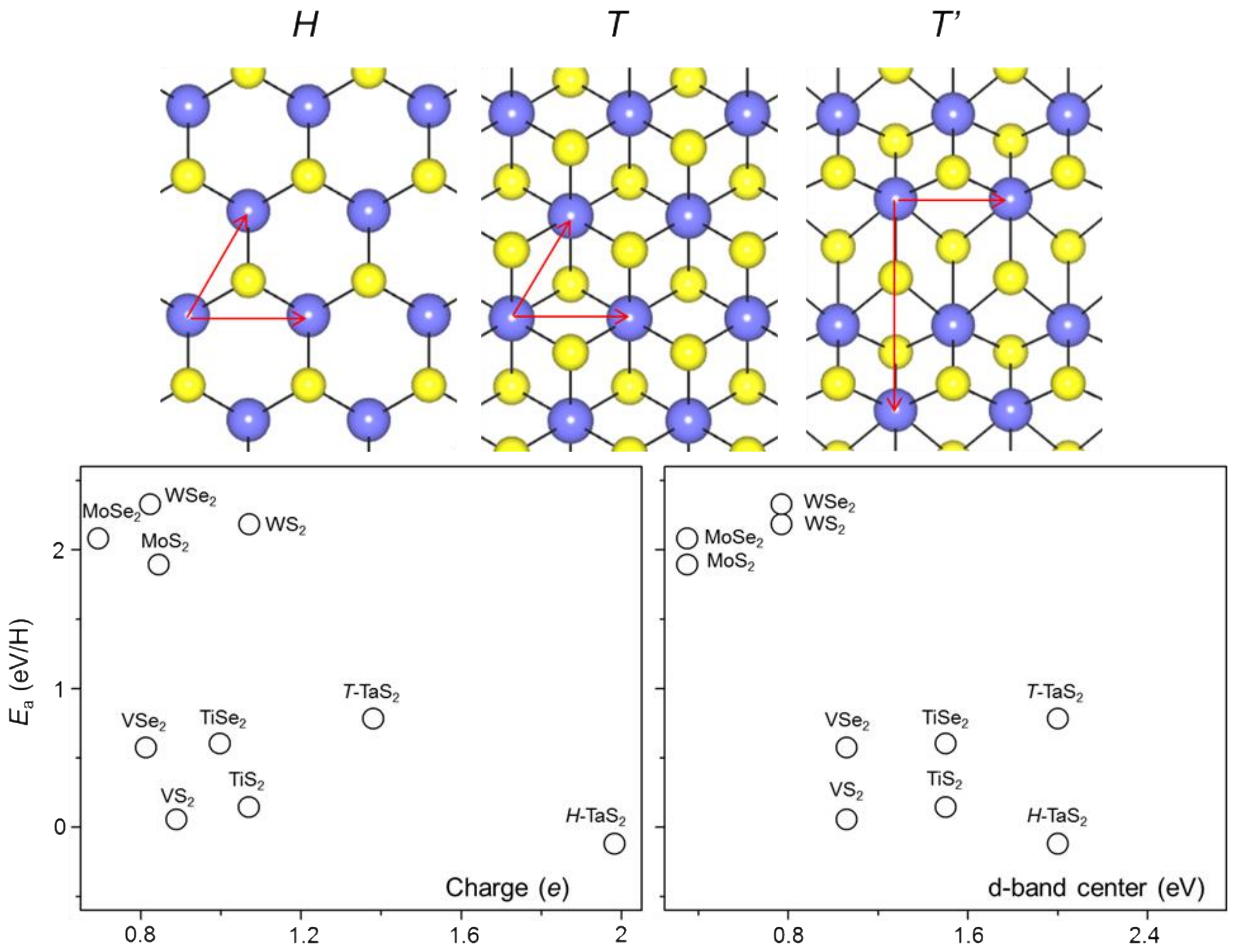

**Figure S2 | Structures of various phases of $MX_2$ and testing of other common descriptors.** Top: the red arrow indicates the primitive cell. Bottom: adsorption energy $E_a$ (Eq. 1, Methods) as a function of the charge on the X atom (left); the *d*-band center of bulk M (right). The *d*-band center values are taken from Ref. [15].

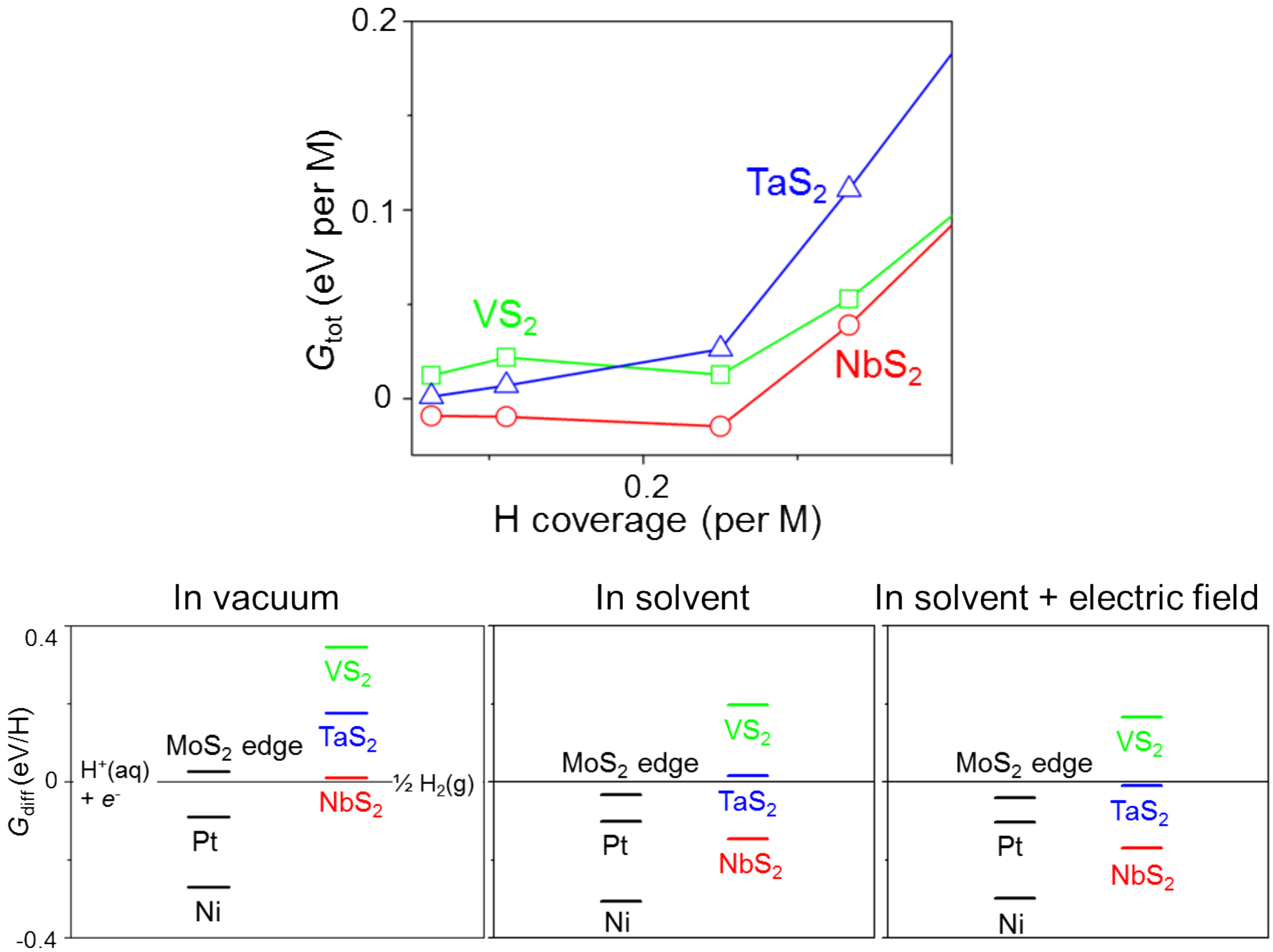

**Figure S3 | Computed $G_{diff}$ and $G_{tot}$ (Eqs. 2 and 3, Methods) for H adsorption on the group-5 $MX_2$ candidates**. Top: Coverage-dependent $G_{tot.}$ Bottom: $G_{diff}$ for the system in vacuum (left), in solvent based on an implicit solvation model (center), and in solvent with a large electric field pointing towards the electrode (right). See Methods for calculation details. For calculation of $G_{diff,}$, results are based on adsorption on a 4x4 supercell; Pt and Ni results assume adsorption on the (111) surface; and the $MoS_2$ edge is modeled using a nanoribbon with edge structure taken from Ref. [8].

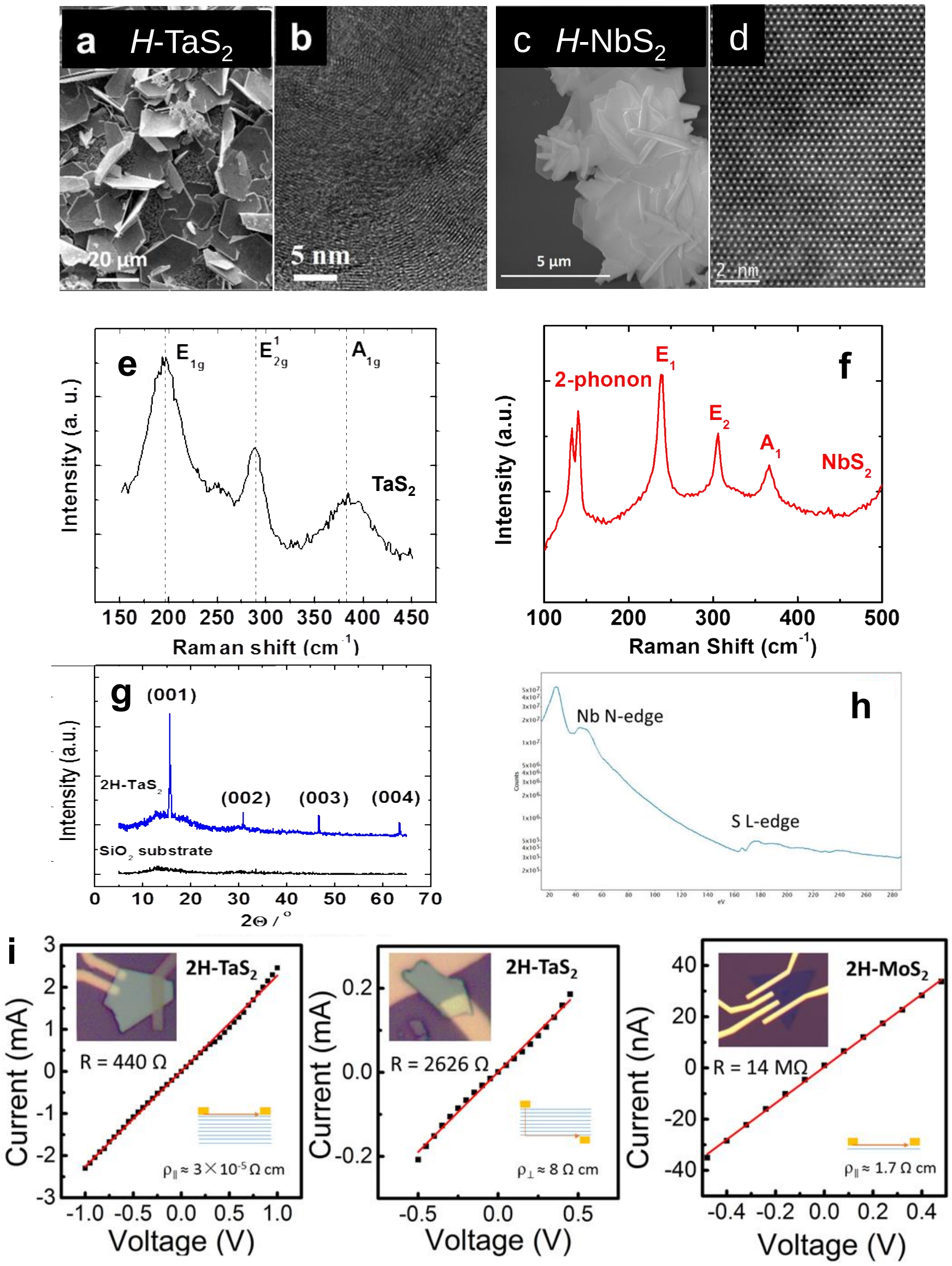

**Figure S4 | Physical characterization of as-grown *H*-$TaS_2$ and *H*-$NbS_2$.** (a-d) SEM (a,c) and TEM (b,d) images of (a-b) *H*-$TaS_2$ and (c-d) *H*-$NbS_2$. (e-f) Raman spectroscopy of (e) *H*-$TaS_2$ and (f) *H*-$NbS_2$. (g) XRD pattern of *H*-$TaS_2$ showing preferred (*00l*) orientation on $SiO_2$/Si substrate. (h) EELS confirming presence of Nb and S in *H*-$NbS_2$. (i) Measurement of the in-plane and out-of-plane resistivity of pre-cycled *H*-$TaS_2$, and in-plane resistivity of *H*-$MoS_2$.

**Table S1.** Comparison of HER activity

| **Sample** | **Catalyst loading ($\mu g/cm^2$)** | **$j_0$ ($A/cm^2$)** | **Tafel slope (mV/decade)** | **\|j\| @ -0.15 V vs. RHE ($mA/cm^2$)** | **\|V\| vs. RHE @ j=-10 $mA/cm^2$ (V)** | **Ref** |
|---|---|---|---|---|---|---|
| **Nano particulate $MoS_2$** | N/A | $1.3\text{-}3.7\times10^{-7}$ | 55-60 | 0.1 | 0.17 | 16 |
| **Particulate $MoS_2$** | 4 | $4.6\times10^{-6}$ | 120 | 0.5 | 0.3 | 17 |
| **Double gyroid $MoS_2$** | 60 | $6.9\times10^{-7}$ | 50 | 1 | 0.28 | 18 |
| **Edge exposed $MoS_2$ film** | 8.5 | $2.2\times 10^{-6}$ | 105-120 | 0.06 | >0.4 | 19 |
| **Edge exposed $MoS_2$ film** | 22 | $1.71\text{-}3.40\times 10^{-6}$ | 115-123 | 0.1 | 0.21 | 20 |
| **30 nm $MoS_2$** | 3400-3900 | $5.0\times 10^{-5}$ | 66 | 10.3 | 0.15 | 21 |
| ***T*- $MoS_2$** | 50 | N/A | 40 | 1 | 0.2 | 11 |
| ***T*- $MoS_2$** | N/A | N/A | 43 | 2 | 0.2 | 22 |
| **$MoS_2$/RGO** | 285 | $5.1\times10^{-6}$ | 41 | 8 | 0.16 | 23 |
| **Defect-Rich $MoS_2$** | 285 | $8.9\times10^{-6}$ | 50 | 3 | 0.2 | 24 |
| **Electrodeposited $MoS_2$** | N/A | N/A | 106 | 2 | >0.4 | 25 |
| **$MoS_2$/CNT-graphene** | 650 | $2.91\times10^{-5}$ | 100 | 2 | 0.28 | 26 |
| **Nanoflakes$WS_2$** | 350 | N/A | 48 | 1 | >0.4 | 27 |
| **$WS_2$/RGO** | 400 | N/A | 58 | 2 | 0.26 | 28 |
| ***T*-$WS_2$** | 6.5 | $1\times10^{-6}$ | 60 | 1 | 0.23 | 29 |
| ***T*-$WS_2$** | 1000±200 | $<10^{-4}$ | 70 | 14 | 0.14 | 30 |
| **$MoS_X$/CNT** | 102 | $33.11\times 10^{-6}$ | 40 | 25 | 0.11 | 31 |
| ***H*-$MoS_2$** | 100 | $3.4\times10^{-6}$ | 120 | 0.08 | >0.4 | This work |
| ***T*-$MoS_2$** | 50 | $2.5\times 10^{-6}$ | 78 | 0.2 | 0.29 | This work |
| ***T*-$TaS_2$** | 80 | $6.4\times 10^{-8}$ | 92 | 0.02 | >0.4 | This work |
| ***H*-$TaS_2$** | 55 | $7.8\times 10^{-5}$ | 37 | 160 | 0.06 | This work |
| ***H*-$NbS_2$** | 10 | $1.3\times10^{-3}$ | 28 | 128 | 0.05 | This work |

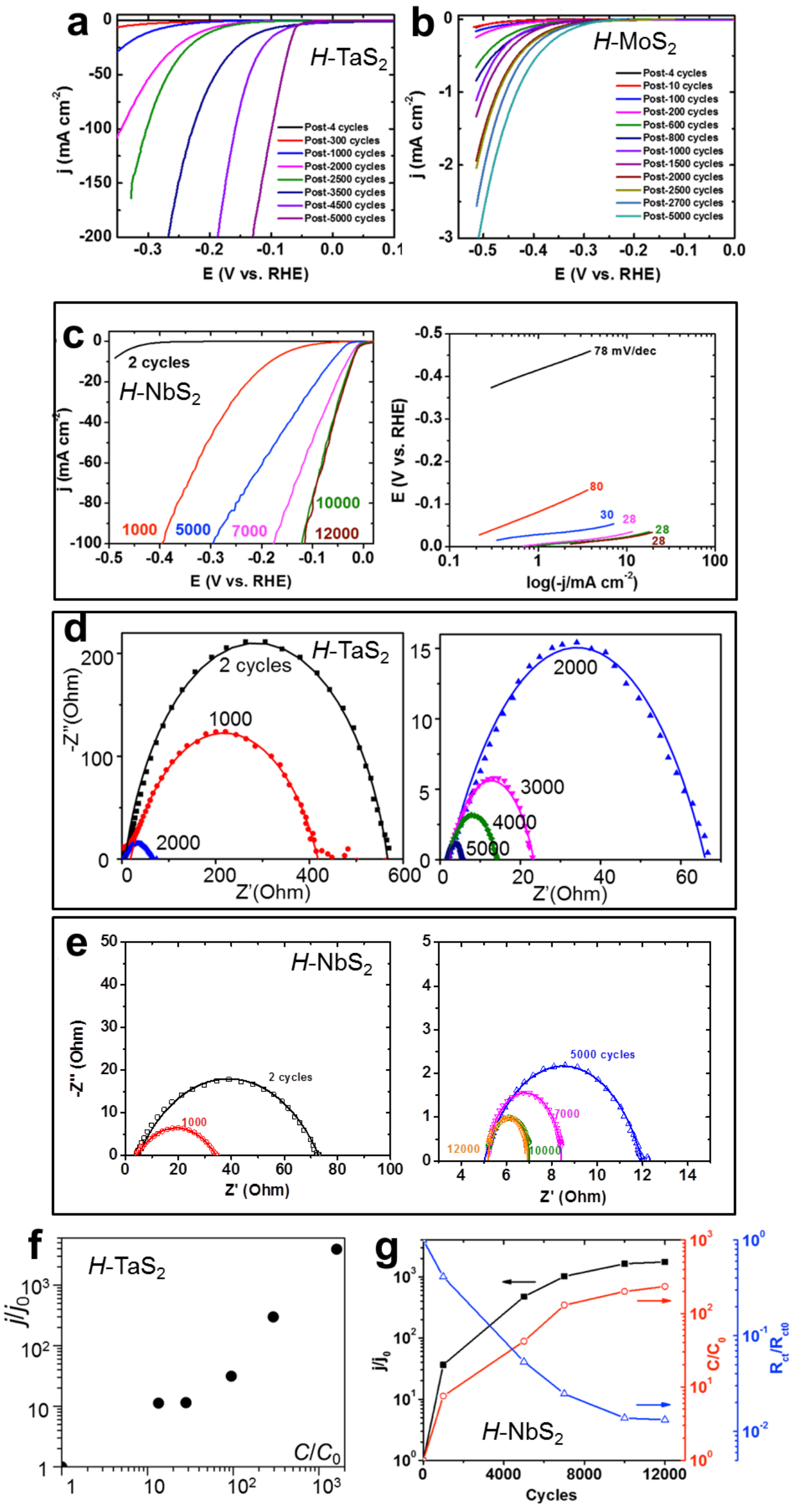

**Figure S5 | Electrochemical evolution recorded periodically during potential cycling.** (a-c): Polarization curves for (a) *H*-$TaS_2$, (b) *H*-$MoS_2$ and (c) *H*-$NbS_2$. The Tafel slope evolution for *H*-$NbS_2$ is also shown in (c). (d-e) EIS for (d) *H*-$TaS_2$ and (e) *H*-$NbS_2$. Solid lines are fits to the equivalent circuit model described in Methods. (f) Relative change of current density $j$ (at -0.1 V vs. RHE) for *H*-$TaS_2$ as a function of the change in the effective double-layer capacitance $C$ for *H*-$TaS_2$. The capacitance is derived from the constant phase element (CPE) in the EIS equivalent circuit. The power of $n$ used in the CPE fits are 0.78, 0.72, 0.76, 0.68, 0.70, 0.72 for 2, 1000, 2000, 3000, 4000 and 5000 cycles, respectively. (g) Cycle-dependent evolution of current, resistivity, and capacitance for *H*-$NbS_2$.

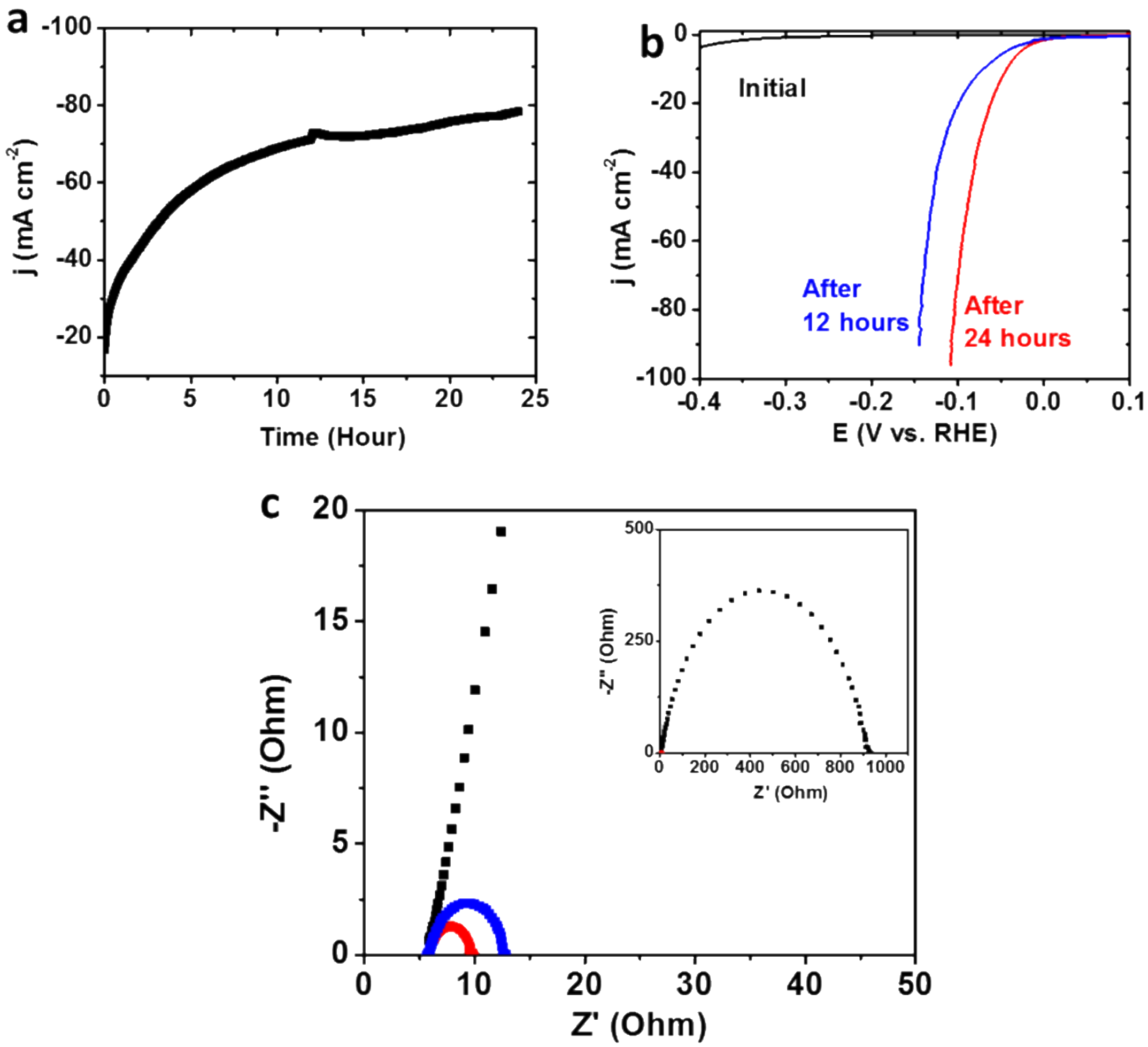

**Figure S6 | HER of *H*-$TaS_2$ under potentiostatic operation**. (a) Chronoamperometric response at constant potential of -0.54 V vs. RHE (not iR corrected). (b) Linear sweep voltammetry of initial and after 12 and 24 hours. (c) Corresponding EIS data.

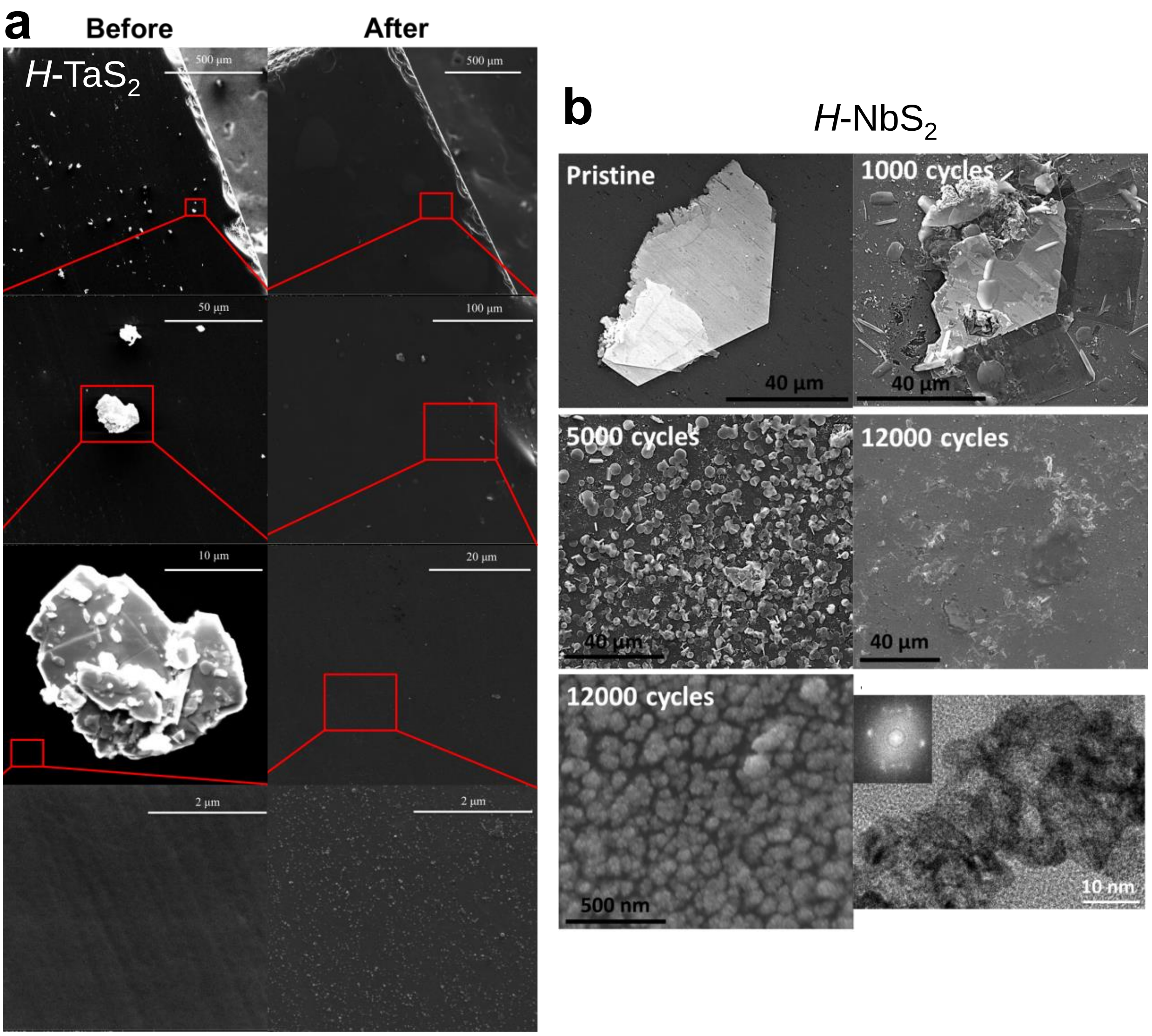

**Figure S7 | Morphological characterization of *H*-TaS$_2$ and *H*-NbS$_2$ upon cycling.** (a) Evolution of *H*-TaS$_2$ from SEM images at different magnifications before (left) and after (right) 5000 cycles. (b) Evolution of H-NbS2 via SEM after different numbers of cycles. The final panel is a TEM image with an FFT inset showing the hexagonal symmetry. Materials are deposited onto a glassy carbon plate electrode for direct comparison of identical regions before and after cycling.

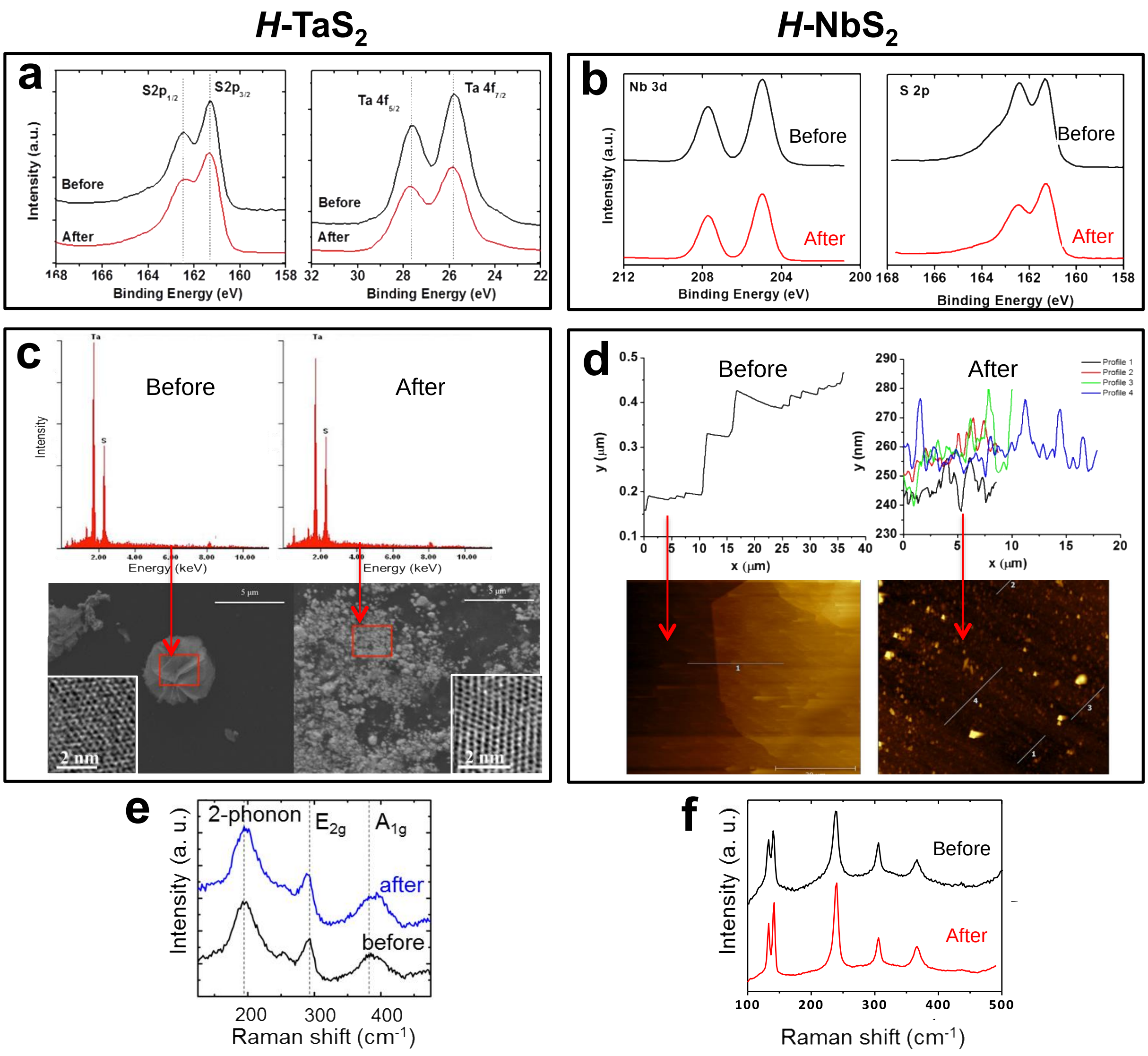

**Figure S8 | Chemical and structural characterization of *H*-$TaS_2$ and *H*-$NbS_2$ on a glassy carbon plate before and after cycling.** (a-b) XPS data for fine scan of S 2p, Ta 4f, and Nb 3d for (a) *H*-$TaS_2$ and (b) *H*-$NbS_2$. (c) EDS characterization of *H*-$TaS_2$ before (left) and after (right) cycling for densely covered regions shown in the SEM images below. Image insets show HRTEM. (d) Thickness profiles for *H*-$NbS_2$ before (left) and after (right) cycling, based on the AFM images below. Four thickness profiles are shown for the cycled sample to confirm thickness uniformity. (e-f) Raman spectroscopy of (a) *H*-$TaS_2$ and (b) *H*-$NbS_2$ before and after cycling. Data for *H*-$TaS_2$ is taken after 5000 cycles; data for *H*-$NbS_2$ is taken after 12000 cycles.

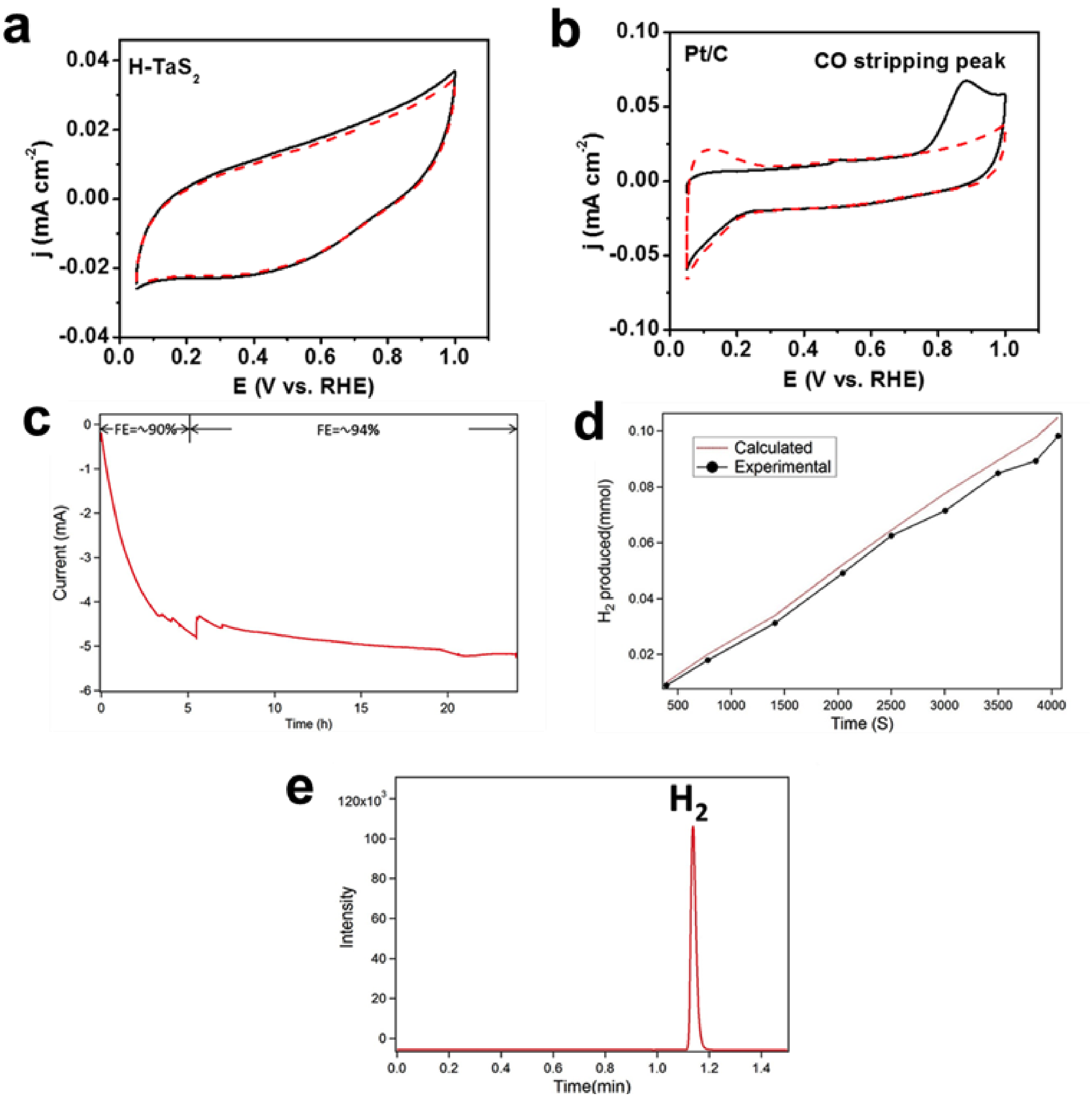

**Figure S9 | Absence of contaminants in *H*-$TaS_2$ and purity of produced $H_2$ gas.** CO stripping voltammetry results on (a) a long-term cycled *H*-$TaS_2$ and (b) a reference Pt/C (Pt loading 0.3 μg $cm^{-2}$) electrode in CO-free solution of 0.5 M $H_2SO_4$. Solid line is CO stripping, dashed line is background cyclic voltammogram collected immediately after CO stripping. The CO is adsorbed at a constant potential of 0.20 V vs. RHE for 5 min. Scan rate is 50 mV/s. (c) Current versus time at -0.4 V vs. RHE, with (d) periodically measured production moles of $H_2$. The red line represents theoretical production moles calculated from the charges passed during electrolysis. (e) Gas chromatography trace of $H_2$ collected from the HER compartment.

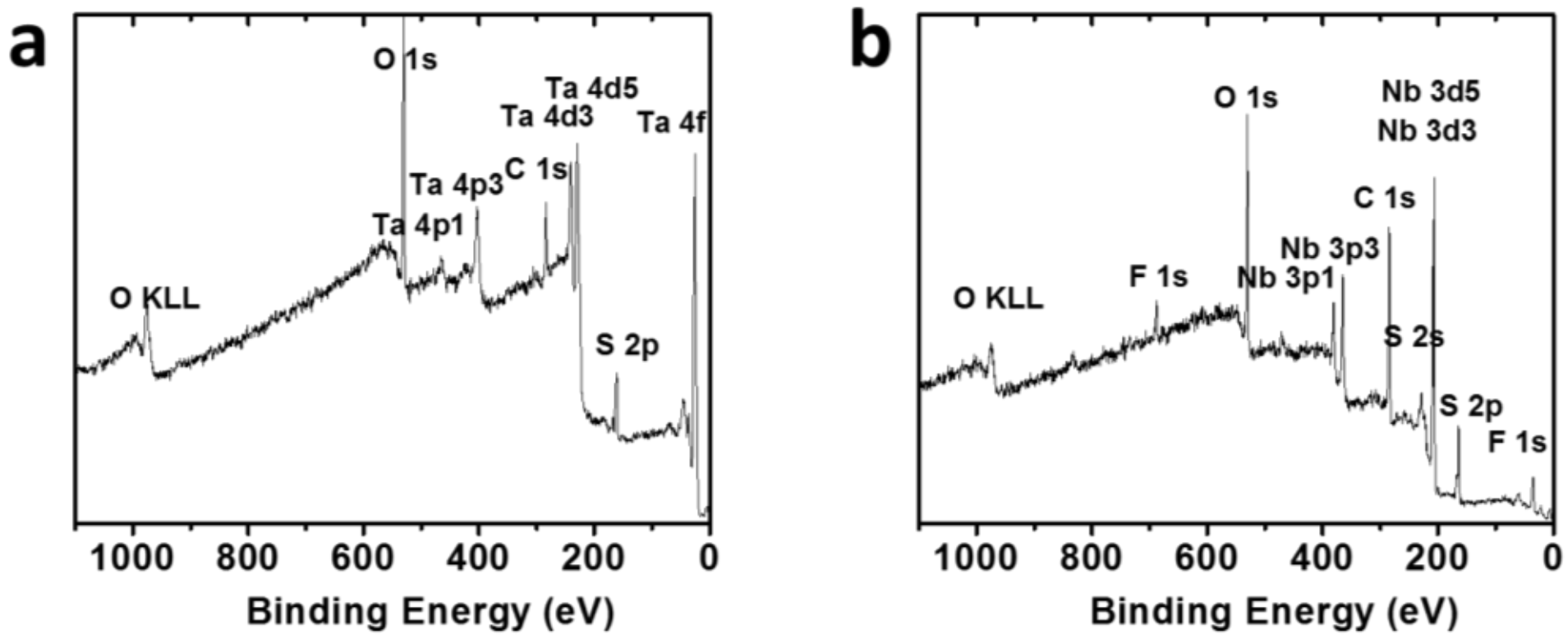

**Figure. S10. XPS survey scan of electrodes after cycling.** (a) H-$TaS_2$ and (b) H-$NbS_2$. The survey scan shows no other metal contaminations on the electrodes or below the detecting limit.

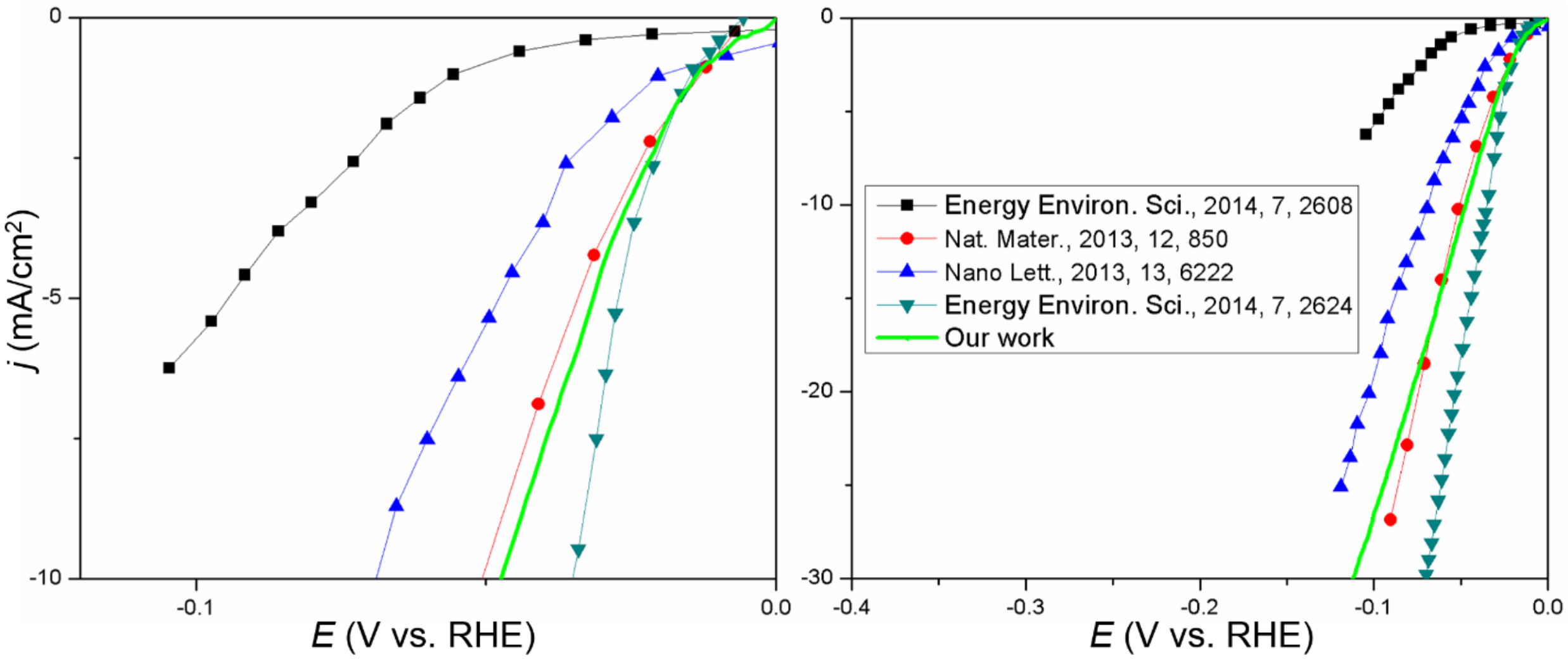

**Figure. S11**. Polarization curves of Pt shown in different scales. The literature results are included for comparison.